\documentclass[english,prl,twocolumn,superscriptaddress,showpacs]{revtex4}
\usepackage[T1]{fontenc}
\usepackage[latin1]{inputenc}
\usepackage{graphicx}

\makeatletter
\usepackage{babel}
\makeatother
\begin{document}

\title{Subcritical crack growth: the microscopic origin of Paris's law}

\author{Andr\'e P. Vieira}
\affiliation{Departamento de Engenharia Metal\'urgica e de
Materiais,
Universidade Federal do Cear\'a, Campus do Pici, 60455-760 Fortaleza,
Cear\'a, Brazil}
\author{Jos\'e S. Andrade Jr.}
\author{Hans J.
Herrmann}
\affiliation{Departamento de F\'{\i}sica, Universidade Federal
do Cear\'a, 60451-970 Fortaleza, Cear\'a, Brazil}
\affiliation{Computational Physics, IfB, ETH-H\"onggerberg,
Schafmattstrasse 6,
8093 Z\"urich, Switzerland}

\begin{abstract}
We investigate the origin of Paris's law, which states that the
velocity of a crack at subcritical load grows like a power law, $da/dt
\sim \left(\Delta K\right)^{m}$, where $\Delta K$ is the stress
intensity factor amplitude.  Starting from a damage accumulation
function proportional to $(\Delta\sigma)^{\gamma}$, $\Delta\sigma$
being the stress amplitude, we show analytically that the asymptotic
exponent $m$ can be expressed as a piecewise-linear function of
the 
%damage accumulation 
exponent $\gamma$, namely, $m=6-2\gamma$
for $\gamma < \gamma_{c}$, and $m=\gamma$ for $\gamma \ge
\gamma_{c}$,
reflecting the existence of a critical value $\gamma_{c}=2$.
%In this way, here we discover the existence of a critical
%value $\gamma_{c}=2$ characterized by a scaling law with a critical
%exponent separating two regimes of different linear functions $m
%(\gamma)$. 
We performed numerical simulations to confirm this result
for finite sizes.  Finally, we introduce bounded
disorder in the breaking thresholds and find that below $\gamma_{c}$
disorder is relevant, i.e., the exponent $m$ is changed, while
above $\gamma_{c}$ disorder is irrelevant.

\pacs{62.20.mm, 46.50.+a, 64.60.av}
% We investigate the origin of Paris's law which states that the
% velocity of a crack at subcritical load grows like a power law, $da/dt
% \sim \left(\Delta K\right)^{m}$, where $\Delta K$ is the stress
% intensity factor amplitude.
% Starting from a damage accumulation function proportional to
% $(\Delta\sigma)^{\gamma}$, $\Delta\sigma$ being the stress amplitude,
% we show analitically that the asymptotic
% exponent $m$ can be expressed as a linear function of the damage
% accumulation exponent $\gamma$. We performed numerical simulations
% to confirm this result for finite sizes. Moreover, we discover the
% existence of a critical value $\gamma_{c}=2$ characterized by a
% scaling law with a critical exponent separating two regimes of
% different functions $m (\gamma)$. Finally, we also study the
% introduction of bounded disorder in the breaking thresholds and find
% that below $\gamma_{c}$ disorder is relevant, i.e., the exponent
% $m$ is changed, while above $\gamma_{c}$ disorder is irrelevant.
\end{abstract}
\maketitle

In 1963 Paris and Erdogan \cite{Paris1963} postulated that under fatigue loading a subcritical crack grows with a velocity that increases with the stress intensity factor, or equivalently the crack length, as a power law with an empirically determined exponent $m$. Numerous experiments showed that this law is valid over at least three orders of magnitude for a very wide spectrum of materials \cite{Lindley1976}. 
%Wei1978, Allen1988a, Mach2007}. 
The Paris law (also known as the
Paris-Erdogan law) had huge implications in engineering since it allowed to predict the residual lifetime of loaded materials quantitatively. Today this law constitutes part of basic knowledge and is taught in elementary courses on mechanics \cite{Suresh1998}. %Schijve1978}. 
Although there
has been attempts to derive the Paris law in terms of geometrical and
crack-tip damage-accumulation models (see e.g. Ref. \cite{Suresh1998} and references
therein), no work has been capable of establishing a firm connection
between the Paris exponent and microscopic parameters.
%\textbf{But nobody since has been able to explain its origin from the microscopic point of view.}
It is the aim of our paper to present analytical and numerical calculations relating the Paris exponent $m$ to the local damage accumulation law. This constitutes a micro-macro derivation of the Paris law. 
%\textbf{Since we showed recently \cite{Kun2007} that the local damage accumulation exponent is identical to the exponent in Basquin's law, which tells how the number of cycles to failure depends on the sub-critical stress, our result finally relates these two important laws on a solid microscopic basis.}

For bodies under cyclic load with stress range $\Delta\sigma$, subcritical fatigue crack growth follows the Paris law \cite{Paris1963},\begin{equation}
\frac{da}{dt}\sim\left(\Delta K\right)^{m}=C\left(\Delta\sigma\sqrt a\right)^{m},
\label{eq:Paris}
\end{equation}
where $a$ is the crack half-length, $\Delta K\sim\Delta\sigma\sqrt{a}$ is the stress-intensity factor range, and $m$ is a material-dependent exponent. Integration over time leads to the Basquin law \cite{Basquin1910},
$%\begin{equation}
 t_{f}\sim \left(\Delta\sigma\right)^{-m}
$, %\label{eq:Basquin}
%\end{equation}
which relates the lifetime $t_{f}$ (or equivalently the number of loading cycles to failure) to the stress amplitude.
Let us note that while the Basquin law applies to high-cycle fatigue
with an exponent $m$ depending on the material structure,
at low-cycle fatigue the corresponding empirical relation,
$t_{f}\sim \left(\Delta\epsilon\right)^{-\lambda}$,
where $\Delta \epsilon$ is the plastic deformation, is called the Coffin-Manson
law and has an exponent $\lambda$ that is remarkably close to
$2$, at least for polycrystalline single-phased metals
\cite{Brechet1992}.
%Sornette1992}.
Recent work \cite{Kun2007} has shown that, in the rupture of fiber-bundle models subject to fatigue damage governed by damage-accumulation functions of the form $\left(\Delta\sigma\right)^{\gamma}$, the Basquin law is verified with an exponent given by the damage-accumulation exponent $\gamma$. This brings the question as to whether it is possible to establish a direct connection between a microscopic damage accumulation form and the Paris law, via the corresponding exponents.
%\textbf{
Here, the term `microscopic' denotes the micrometer rather than the atomic
(nanometer) length scale.
%}

\begin{figure}
\includegraphics[width=0.6\columnwidth]{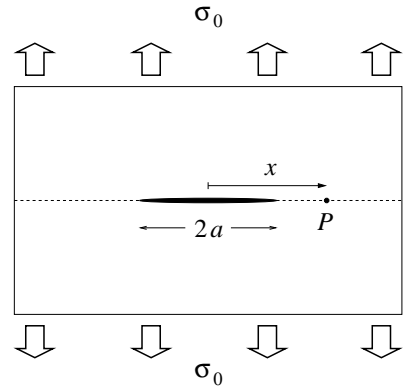}
\caption{\label{fig:crack}A very thin crack of length $2a$, propagating
along the dashed line in a planar medium subject to a stress
$\sigma_{0}$ at infinity. Point $P$ is at a distance $x$ from the crack center.}
\end{figure}

In order to address this question, we consider a linear crack of length $2a$, in a two-dimensional medium
subject to a transverse external stress $\sigma_{0}$ exerted very
far from the crack, as depicted in Fig. \ref{fig:crack}. We model
the medium as composed of small elements connected by stiff elastic springs,
with separation $\delta r$. In the continuum limit, and
within linear elasticity theory, the transverse stress at a point
along the crack line, a distance $x$ from the midpoint of the crack,
is given by \cite{Marder2000}\begin{equation}
\sigma\left(x;a\right)=\sigma_{0}\frac{x}{\sqrt{x-a}\sqrt{x+a}}.\label{eq:cs}\end{equation}
Close to the crack tip ($x\simeq a$), the stress diverges as
$\sigma\left(x;a\right)={K}/{\sqrt{2\pi\left(x-a\right)}}$,
%\begin{equation}
%\sigma\left(x;a\right)=\frac{K}{\sqrt{2\pi\left(x-a\right)}},
%label{eq:stress}\end{equation}
defining the stress intensity factor 
$%\begin{equation}
K=\sigma_{0}\sqrt{\pi a}.
$%\end{equation}

We assume that the medium is under cyclic load, with an external stress
varying between $\sigma_{\mathrm{min}}$ and $\sigma_{\mathrm{max}}$,
leading to a stress-intensity-factor range $\Delta K=\left(\sigma_{\mathrm{max}}-\sigma_{\mathrm{min}}\right)\sqrt{\pi a}$.
We further assume that fatigue damage is the sole factor driving crack
growth, which happens only along the crack line \cite{note1}.
Specifically, the
half-length of the crack increases by $\delta r$ when the accumulated
damage at the crack tip reaches a threshold value $F_{\mathrm{thr}}$.
Damage increments are assumed to be given by \begin{equation}
\delta F\left(x;a\right)=f_{0}\delta t\left(a\right)\left[\Delta\sigma\left(x;a\right)\right]^{\gamma},\label{eq:df}\end{equation}
where $\delta t\left(a\right)$ is the number of cycles during which
the crack has length $2a$, $f_{0}$ is a constant related to the
time scale, 
%$\gamma$ is the Basquin-law exponent \cite{Kun2007},
and $\Delta\sigma\left(x;a\right)$ is calculated from Eq. (\ref{eq:cs}) with $\sigma_0$ varying
between $\sigma_{\mathrm{min}}$ and $\sigma_{\mathrm{max}}$.
The heuristic \cite{note2}
power-law dependence of the damage increment can be justified by
invoking concepts of self-similarity \cite{Barenblatt1983}, and $\gamma$ can
be seen as a stress-amplification exponent \cite{Sornette2006}.

The accumulated damage at point $x$ when the crack has length
$2a$ is given by \begin{equation}
F\left(x;a\right)=F\left(x;a-\delta r\right)+\delta F\left(x;a\right).\label{eq:fxa}\end{equation}
Since the half-length of the crack increases from $a-\delta r$ to $a$
when the accumulated damage $F\left(a+\delta r;a\right)$ reaches $F_{\mathrm{thr}}$,
it follows from Eqs. (\ref{eq:df}) and (\ref{eq:fxa}) that\begin{equation}
\delta t\left(a\right)=\frac{F_{\mathrm{thr}}-F\left(a+\delta r;a-\delta r\right)}{f_{0}\left[ \Delta\sigma\left(a+\delta r;a\right)\right] ^{\gamma}}.\label{eq:dt}\end{equation}
Here, in order to avoid the appearance of infinities, we assume that
the spring attached to an element at position $x$ experiences a stress
given by $\sigma\left(x+\delta r;a\right)$. To a first approximation,
this is consistent with the fact that linear elasticity theory must
break down in the immediate vicinity of the crack tip, giving rise
to a fracture process zone or plastic zone \cite{Suresh1998,Alava2006}.

\begin{figure}
\includegraphics[width=0.8\columnwidth]{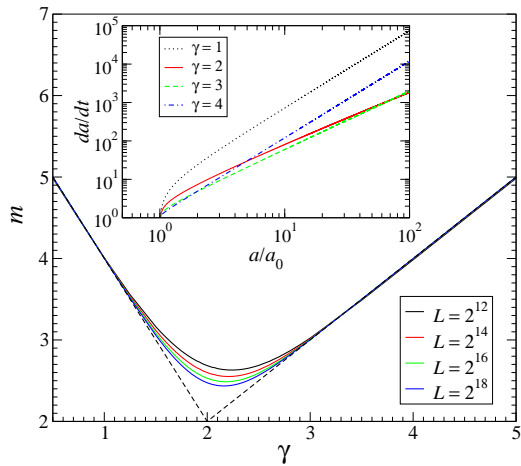}
\caption{\label{fig:bvsg} (Color online)
Inset: Log-log plot of the time derivative of the crack
half-length $a$, as a function of $a/a_{0}$,
for various values of the damage exponent $\gamma$. In each case,
the time scale was adjusted so that $da/dt$ approaches unity as $a$
approaches the initial value $a_{0}=100\delta r$. Main panel: Dependence of the Paris exponent $m$
on $\gamma$, for different values of the system size $L$. The dashed curve corresponds to Eq. (\ref{eq:bvsg}).}
\end{figure}
\begin{figure}
\includegraphics[width=0.8\columnwidth]{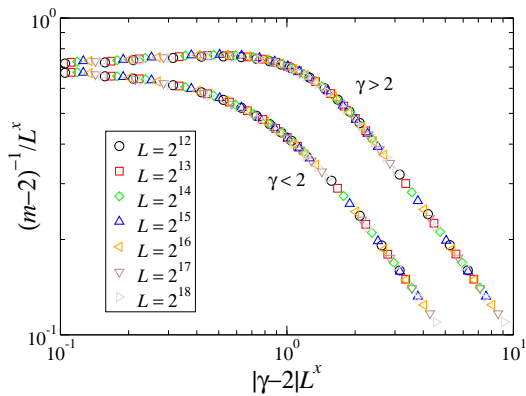}

\caption{\label{fig:fss} (Color online) Scaling plot of $m$ versus $\gamma$ for different
values of the system size $L$, showing conformance with
Eq. (\ref{eq:fss}). The best data collapse is obtained with $x=0.089$.}
\end{figure}

Numerical iteration of the above equations reveals a time dependence
of the crack length $2a$ which, for large values of $a$, reproduces
the Paris law, Eq. (\ref{eq:Paris}),
with a $\gamma$-dependent exponent $m$, as shown in the inset in Fig. \ref{fig:bvsg}.
Notice that $m$ (as determined from the slopes of the log-log
plots) seems to reach a minimum value for $\gamma\simeq2$.
This can be explained as a result of the competition between damage accumulation
mostly around the crack tip, which happens for $\gamma\gg 1$, and uniform damage along the whole crack line, which dominates as $\gamma$ approaches zero.
The existence of this minimum is confirmed by calculations of $m$ as a function of
$\gamma$, obtained by power-law fits of $da/dt$ versus $a$, using
various system sizes $L$. 
% The fits were extracted
% from points whose abscissa values lie between $90\%$ and $100\%$
% of $L$. 
As shown in the main panel in Fig. \ref{fig:bvsg}, there are strong finite-size
effects around $\gamma_{c}=2$. As $L\rightarrow\infty$, the minimum
in each curve shifts slowly towards $m=2$ at $\gamma_{c}=2$ , suggesting
an asymptotic form\begin{equation}
m=\left\{ \begin{array}{lr}
6-2\gamma, & \mbox{for }\gamma<\gamma_{c}\\
\gamma, & \mbox{for }\gamma>\gamma_{c}\end{array}\right..\label{eq:bvsg}\end{equation}
This can be checked by a finite-size scaling hypothesis,\begin{equation}
\frac{1}{m\left(L\right)-2}=\left\{ \begin{array}{lr}
L^{x}\mathcal{F}_{-}\left(\left|\gamma-2\right|L^{x}\right), & \mbox{for }\gamma<\gamma_{c}\\
L^{x}\mathcal{F}_{+}\left(\left|\gamma-2\right|L^{x}\right), & \mbox{for }\gamma>\gamma_{c}\end{array}\right.,\label{eq:fss}\end{equation}
which, as shown in Fig. \ref{fig:fss}, is nicely fulfilled by our
numerical data with $x=0.089$. Both scaling functions
behave as
$%\begin{equation}
\mathcal{F}_{\pm}\left(u\right)\sim u^{-1}%, \qquad u\gg1,
$%\end{equation}
, for $u\gg 1$,
in agreement with the suggestion that, in the continuum limit ($L\gg\delta r$),
$m$ should be a linear function of $\gamma$ for both $\gamma<\gamma_{c}$
and $\gamma>\gamma_{c}$. We have therefore found evidence that $\gamma = \gamma_{c}$ is a critical point,
% with a critical exponent $x$.
associated with the divergence of the stress integral along the crack line.
%associated with the divergence of the length scale above which the Paris law is verified.
%\textbf{
Our conclusion that $m\geq 2$ is fully compatible with values reported for various materials throughout the literature.
%}

The prediction of Eq. (\ref{eq:bvsg}) is confirmed by an analytical treatment
of Eqs. (\ref{eq:df})-(\ref{eq:dt}). From
the time $\delta t\left(a_{0}\right)$ during which the crack length is $2a_{0}$,
we can calculate the instantaneous damage $F\left(a_{0}+2\delta r;a_{0}\right)$ at the next crack tip position $a_{0}+2\delta r$, and then recursively determine the damage fraction at the successive crack tip positions, 
$G_{n}\equiv F\left(a+\delta r;a-\delta r\right)/F_{\mathrm{thr}}$,
%$G_{n}\equiv F\left(a_{0}+(n+1)\delta r;a_{0}+(n-1)\delta r\right)/F_{\mathrm{thr}}$, 
from 
$\delta t\left(a-\delta r\right)$
%$\delta t\left(a_{0}+(n-1)\delta r\right)$
and
$F\left(a+\delta r;a-2\delta r\right)$,
%$F\left(a_{0}+(n+1)\delta r;a_{0}+(n-2)\delta r\right)$.
with $a=a_{0}+n\delta r$.
As a result, it is possible to write the recurrence relation \begin{equation}
 G_{n}+\sum_{k=1}^{n}g_{n,k}G_{k-1}=\sum_{k=1}^{n}g_{n,k}, \label{eq:rr}
\end{equation}
with $G_{0}=0$ and
\begin{equation}
 g_{n,k}=\left[\frac
{\Delta\sigma\left(a_{0}+(n+1)\delta r;a_{0}+(k-1)\delta r\right)}
{\Delta\sigma\left(a_{0}+k\delta r;a_{0}+(k-1)\delta r\right)}
\right]^\gamma . \label{eq:gnk}
\end{equation}
Notice that $g_{n,k}$ is determined by the ratio between the stress amplitudes at position $n\delta r$ and at the crack tip, when the crack has grown by a distance $2k\delta r$.
From Eqs. (\ref{eq:dt}) and (\ref{eq:cs}), it follows that, for $a=a_{0}+n\delta r$, with $n\gg 1$,
\begin{equation}
 \frac{da}{dt}=\frac{\delta r}{\delta t(a)}\sim 
\frac {n^{\frac{\gamma}{2}}}{1-G_{n}}, \label{eq:dadtn}
\end{equation}
so that the scaling behavior of the crack growth rate depends on the scaling behavior of $G_n$. This can be investigated by looking at Eqs. (\ref{eq:rr}) and (\ref{eq:gnk}), from which, if $\delta r\ll a_{0}\ll n\delta r$, we obtain the asymptotic forms
\begin{equation}
 g_{n,k}\simeq \left\{ \begin{array}{ll}
\left({2\delta r}/{a_{0}}\right)^{\frac{1}{2}\gamma}, & \mbox{for }k\ll a_{0}/\delta r;\\
2^{\frac{\gamma}{2}}k^{-\frac{1}{2}\gamma}, & \mbox{for }a_{0}/\delta r\ll k\ll n;\\
\left(n-k+2\right)^{-\frac{1}{2}\gamma}, & \mbox{for }k\lesssim n .
\end{array}\right.
\end{equation}
It is also easy to show that $g_{n,k}$ has a single minimum at $k_{\mathrm{min}}\simeq n/\sqrt  3$, so that, as $n\rightarrow\infty$, the sum on the right-hand side of 
Eq. (\ref{eq:rr}) has a power-law divergence $n^{1-\frac{1}{2}\gamma}$ for $\gamma <2$, while it converges to a finite value $C_{\gamma}=\zeta(\frac{1}{2}\gamma)-1$ for $\gamma >2$, where $\zeta(x)$ is the Riemann zeta function. It follows that $G_{n}$ approaches $C_{\gamma}/(1+C_{\gamma})<1$ for $\gamma>2$, and from Eq. (\ref{eq:dadtn}) we immediately see that $m=\gamma$. On the other hand, numerical calculations show that, for $\gamma <2$, $G_{n}$ asymptotically approaches unity; although we were not able to derive an analytic expression, it can be readily checked numerically that in this case
\begin{equation}
 1-G_{n}\sim n^{-3\left(1-\frac{1}{2}\gamma\right)},
\end{equation}
leading, when combined with Eq. (\ref{eq:dadtn}), to $m=6-2\gamma$.

For $\gamma>\gamma_{c}=2$, in the light of Basquin's law, the prediction $m=\gamma$ is compatible with the result obtained in Ref. \cite{Kun2007}, stating that the Basquin-law exponent is given by
the damage exponent $\gamma$. This is no longer valid for $\gamma<\gamma_{c}$. However,  the models studied in Ref. \cite{Kun2007} involve randomness in both fatigue and stress thresholds as additional ingredients.

In order to investigate the effects of disorder on the Paris exponent, we
introduce a distribution of values of the fatigue thresholds with lower (upper)
cutoff $F_{1}$ ($F_{2}$), so that
each lattice point has a local threshold $F_{1}\leq F_{\mathrm{thr}}(x)\leq F_{2}$.
An obvious consequence of the disorder is that, depending on the disorder strength and on the damage exponent $\gamma$, it is possible that a point far from the crack tip reaches its local fatigue threshold
\emph{before} a point closer to the crack tip. This leads to the occurrence of rupture avalanches, in much the same way as in fiber-bundle models (see e.g. Ref. \cite{Pradhan2005}).

Conditions for the appearance of avalanches can be derived from the analytical approach presented above. Avalanches will occur if, for some crack length $2a$, the local threshold at position $a+\delta r$ is less than the accumulated damage at that position when the crack had length $2(a-\delta r)$,
i.e. $%\begin{equation}
 F_{\mathrm{thr}}\left(a+\delta r\right) < F\left(a+\delta r;a-\delta r\right)%.
$. %\end{equation}
As the fatigue thresholds are no longer the same for all points, Eq. (\ref{eq:rr}) ceases to be valid. However, $F\left(a+\delta r;a-\delta r\right)$ can still be written as a linear combination of the thresholds of the points closer to the crack center, with coefficients related to the stress ratios $g_{n,k}$. In the absence of any previous avalanches, and for $\gamma > \gamma_{c}=2$, the value of $F\left(a+\delta r;a-\delta r\right)$ is limited by $F_{2}$ times the asymptotic value of $G_{n}$,
i.e. $%\begin{equation}
 F\left(a+\delta r;a-\delta r\right)\leq F_{2}{C_{\gamma}}/\left(1+C_{\gamma}\right)%.
$. %\end{equation}
Since $F_{1}\leq F_{\mathrm{thr}}\left(a+\delta r\right)$, we conclude that avalanches will occur if
\begin{equation}
 F_{1}-F_{2}\frac{C_{\gamma}}{1+C_{\gamma}}\leq 0\quad\Rightarrow\quad 
\frac{F_{1}}{F_{2}}\leq \frac{C_{\gamma}}{1+C_{\gamma}}. \label{eq:fc}
\end{equation}
On the other hand, for $\gamma <\gamma_{c}$, 
%this argument indicates that 
any finite amount of disorder leads to the occurrence of avalanches.

%  \begin{figure}
%   \includegraphics[width=0.8\columnwidth]{fig4}
%  \caption{\label{fig:bf34} (Color online) Plots of the deviation from unity of the ratio $B(f)/B(1)$ between the coefficient of Eq. (\ref{eq:ipl}) in the presence and in the absence of disorder, as a function of the ratio between the lower and upper distribution cutoffs.}
%  \end{figure}
Confirmation of this prediction, as well as further information on the effects of disorder, can be obtained from numerical calculations. In order to analyze the results of those calculations, we integrate the Paris law to obtain
\begin{equation}
 1-\left(a_{0}/a\right)^{\frac{1}{2}m-1}=Bt, \label{eq:ipl}
\end{equation}
where $B$ is a $m$-dependent constant related to the inverse rupture time. Introducing fatigue thresholds uniformly distributed between $F_1$ and $F_2$,
%\begin{equation}
% P\left(F_{\mathrm{thr}}\right)=\frac{1}{F_{2}-F_{1}}
%\theta\left(F_{2}-F_{\mathrm{thr}}\right)\theta\left(F_{\mathrm{thr}}-F_{1}\right),
%\end{equation}
numerical iteration of Eqs. (\ref{eq:cs}) and (\ref{eq:df}) --- modified to acomodate local thresholds --- shows that, for $\gamma >\gamma_{c}$, plots of
$1-\left(a_{0}/a\right)^{\frac{1}{2}m-1}$ for large times (not shown) remain straight lines, with the same value of $m=\gamma$ as in the absence of disorder; however, for sufficiently small values of $f=F_{1}/F_{2}$,
%with a fixed average threshold $(F_{1}+F_{2})/2$, 
the coefficient $B$ becomes disorder-dependent.
% as shown  for $\gamma=3$ and $\gamma=4$ in Fig. \ref{fig:bf34}. The values $f_{c}$ of $f$ at which the plots start to deviate from the horizontal dotted line are compatible with the predictions of Eq. (\ref{eq:fc}), namely $f_{c}\simeq 0.62$ and $f_{c}\simeq 0.39$ for $\gamma=3$ and $\gamma=4$, respectively. Indeed, we find $1-B(f)/B(1)\sim\left(f_c-f\right)^{\alpha}$ for $f<f_{c}$, with the exponent $\alpha$ around $2.5$, but we cannot exclude that the
% exponent might depend on $\gamma$.

\begin{figure}
 \includegraphics[width=0.9\columnwidth]{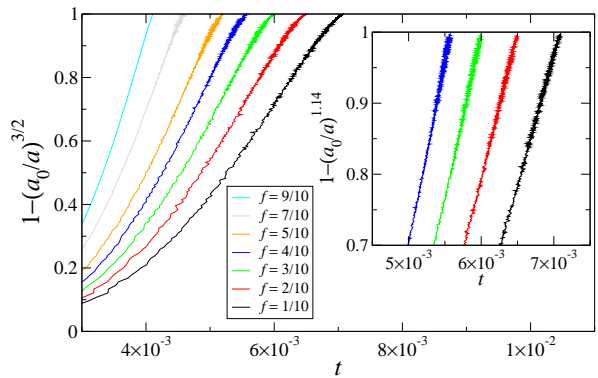}
\caption{\label{fig:avst} (Color online) Rescaled plots of the crack length for $\gamma=\frac{1}{2}$ and various ratios $f=F_{1}/F_{2}$ between the minimum and maximum values of the fatigue thresholds. Main panel: rescaling as given by Eq. (\ref{eq:ipl}). Inset: rescaling with an exponent $1.14\neq \frac{1}{2}m -1$.
All curves were obtained for system size $L=2^{15}$ and averaged over $500$ disorder realizations.}
\end{figure}
For $\gamma <\gamma_{c}$, we confirm the occurrence of avalanches for any amount of disorder. Also, as shown in Fig. \ref{fig:avst} (main panel), the prediction of Eq. (\ref{eq:ipl}) is no longer verified, indicating that the Paris exponent $m$ is actually modified by the introduction of disorder 
\footnote{
This distinction between the effects of the disorder for $\gamma<\gamma_{c}$ and $\gamma>\gamma_{c}$ is reminiscent of the Harris criterion for the relevance of disorder on the critical behavior of ferromagnetic models
[A. B. Harris, J. Phys. C \textrm{7}, 1671 (1974)], 
according to which disorder may change the critical exponents of the system only if the uniform specific heat exponent $\alpha$ is positive.
}. 
The new exponent $m^{\prime}$, which recovers the linear behavior predicted by Eq. (\ref{eq:ipl}), is almost independent of the ratio $f$, for sufficiently strong disorder, as also shown in Fig. \ref{fig:avst} (inset), and we have checked that in this limit the ratio $(m^{\prime}-2)/(m-2)$ is independent of $\gamma$, being given approximately by $0.76$.

We have been able to derive analytically and confirm numerically the Paris law and found that its exponent $m$ is a function of the damage exponent $\gamma$ describing the microscopic damage accumulation law. To our surprise we discovered that $\gamma_{c}=2$ is a critical point characterized by a scaling law and a critical exponent which separates two regimes with different linear functions $m (\gamma)$. In addition we also studied the role of disorder and found again that $\gamma_{c}=2$ plays a special role: disorder is relevant below it and irrelevant above it.
Our results can have far-reaching consequences in the understanding and
control of subcritical crack propagation. On one hand the discovered
relation between the damage and the Paris exponents, 
which in principle could be checked experimentally,
could help to predict lifetimes of samples
by studying the velocity of small cracks. As we found, the value
$\gamma_c=2$ is very special. For example, to avoid the influence of
disorder one must try to stay above $\gamma_c$. Finally, it is important
to mention that the exponent $\gamma$ of the damage accumulation law is
material dependent and could therefore play a central role in the
engineering design to increase the robustness and optimize the mechanical
performance of the system. 
%\textbf{
The derivation of power-law damage-accumulation functions from atomistic processes
would be of much interest, and represents a big challenge for future work.
%}

We would like to thank Stefano Zapperi for helpful discussions and
the Brazilian agencies CNPq, CAPES, FINEP and FUNCAP for financial
support. HJH thanks the Max Planck prize.

%\bibliographystyle{apsrev}
%\bibliography{fraturas}

\begin{thebibliography}{20}
\expandafter\ifx\csname natexlab\endcsname\relax\def\natexlab#1{#1}\fi
\expandafter\ifx\csname bibnamefont\endcsname\relax
  \def\bibnamefont#1{#1}\fi
\expandafter\ifx\csname bibfnamefont\endcsname\relax
  \def\bibfnamefont#1{#1}\fi
\expandafter\ifx\csname citenamefont\endcsname\relax
  \def\citenamefont#1{#1}\fi
\expandafter\ifx\csname url\endcsname\relax
  \def\url#1{\texttt{#1}}\fi
\expandafter\ifx\csname urlprefix\endcsname\relax\def\urlprefix{URL }\fi
\providecommand{\bibinfo}[2]{#2}
\providecommand{\eprint}[2][]{\url{#2}}

\bibitem[{\citenamefont{Paris and Erdogan}(1963)}]{Paris1963}
\bibinfo{author}{\bibfnamefont{P.~C.} \bibnamefont{Paris}} \bibnamefont{and}
  \bibinfo{author}{\bibfnamefont{F.}~\bibnamefont{Erdogan}},
  \bibinfo{journal}{J. Basic Eng.} \textbf{\bibinfo{volume}{85}},
  \bibinfo{pages}{528} (\bibinfo{year}{1963}).

\bibitem[{\citenamefont{Lindley et~al.}(1976)\citenamefont{Lindley, Richards,
  and Ritchie}}]{Lindley1976}
\bibinfo{author}{\bibfnamefont{T.~C.} \bibnamefont{Lindley}},
  \bibinfo{author}{\bibfnamefont{C.~E.} \bibnamefont{Richards}},
  \bibnamefont{and} \bibinfo{author}{\bibfnamefont{R.~O.}
  \bibnamefont{Ritchie}}, \bibinfo{journal}{Metall. Met. Form.}
  \textbf{\bibinfo{volume}{43}}, \bibinfo{pages}{268} (\bibinfo{year}{1976});
%
%\bibitem[{\citenamefont{Wei}(1978)}]{Wei1978}
\bibinfo{author}{\bibfnamefont{R.~P.} \bibnamefont{Wei}}, \bibinfo{journal}{J.
  Eng. Mater. Tech.} \textbf{\bibinfo{volume}{100}}, \bibinfo{pages}{113}
  (\bibinfo{year}{1978});
%
%\bibitem[{\citenamefont{Allen et~al.}(1988)\citenamefont{Allen, Booth, and Jutla}}]{Allen1988a}
\bibinfo{author}{\bibfnamefont{R.~J.} \bibnamefont{Allen}},
  \bibinfo{author}{\bibfnamefont{G.~S.} \bibnamefont{Booth}}, \bibnamefont{and}
  \bibinfo{author}{\bibfnamefont{T.}~\bibnamefont{Jutla}},
  \bibinfo{journal}{Fatigue Fract. Eng. Mater. Struct.}
  \textbf{\bibinfo{volume}{11}}, \bibinfo{pages}{45} (\bibinfo{year}{1988});
\bibinfo{journal}{ibid.} \textbf{\bibinfo{volume}{11}}, \bibinfo{pages}{71}
  (\bibinfo{year}{1988}{\natexlab{b}});
%
%\bibitem[{\citenamefont{Mach et~al.}(2007)\citenamefont{Mach, Hale, Denny, and Nelson}}]{Mach2007}
\bibinfo{author}{\bibfnamefont{K.~J.} \bibnamefont{Mach}}
  \bibinfo{author}{\bibfnamefont{\textit{et al.}}},
  %\bibinfo{author}{\bibfnamefont{B.~B.} \bibnamefont{Hale}},
  %\bibinfo{author}{\bibfnamefont{M.~W.} \bibnamefont{Denny}}, \bibnamefont{and}
  %\bibinfo{author}{\bibfnamefont{D.~V.} \bibnamefont{Nelson}},
  \bibinfo{journal}{J. Exp. Biology} \textbf{\bibinfo{volume}{210}},
  \bibinfo{pages}{2231} (\bibinfo{year}{2007}).

\bibitem[{\citenamefont{Suresh}(1998)}]{Suresh1998}
\bibinfo{author}{\bibfnamefont{S.}~\bibnamefont{Suresh}},
  \emph{\bibinfo{title}{Fatigue of materials}} (\bibinfo{publisher}{Cambridge
  University Press}, \bibinfo{year}{1998}), \bibinfo{edition}{2nd} ed;
%
%\bibitem[{\citenamefont{Schijve}(1978)}]{Schijve1978}
\bibinfo{author}{\bibfnamefont{J.}~\bibnamefont{Schijve}},
  \bibinfo{journal}{Eng. Fract. Mech.} \textbf{\bibinfo{volume}{11}},
  \bibinfo{pages}{169} (\bibinfo{year}{1978}).

\bibitem[{\citenamefont{Basquin}(1910)}]{Basquin1910}
\bibinfo{author}{\bibfnamefont{O.~H.} \bibnamefont{Basquin}},
  \bibinfo{journal}{Proceedings of the American Society for Testing and
  Materials} \textbf{\bibinfo{volume}{10}}, \bibinfo{pages}{625}
  (\bibinfo{year}{1910}).

\bibitem[{\citenamefont{Brechet et~al.}(1992)\citenamefont{Brechet, Magnin, and
  Sornette}}]{Brechet1992}
\bibinfo{author}{\bibfnamefont{Y.}~\bibnamefont{Brechet}},
  \bibinfo{author}{\bibfnamefont{T.}~\bibnamefont{Magnin}}, \bibnamefont{and}
  \bibinfo{author}{\bibfnamefont{D.}~\bibnamefont{Sornette}},
  \bibinfo{journal}{Acta Metall. Mater.} \textbf{\bibinfo{volume}{40}},
  \bibinfo{pages}{2281} (\bibinfo{year}{1992});
%
%\bibitem[{\citenamefont{Sornette et~al.}(1992)\citenamefont{Sornette, Magnin,and Brechet}}]{Sornette1992}
\bibinfo{author}{\bibfnamefont{D.}~\bibnamefont{Sornette}},
  \bibinfo{author}{\bibfnamefont{T.}~\bibnamefont{Magnin}}, \bibnamefont{and}
  \bibinfo{author}{\bibfnamefont{Y.}~\bibnamefont{Brechet}},
  \bibinfo{journal}{Europhys. Lett.} \textbf{\bibinfo{volume}{20}},
  \bibinfo{pages}{433} (\bibinfo{year}{1992}).

\bibitem[{\citenamefont{Kun et~al.}(2007)\citenamefont{Kun, Costa, {Costa
  Filho}, {Andrade Jr.}, Soares, Zapperi, and Herrmann}}]{Kun2007}
\bibinfo{author}{\bibfnamefont{F.}~\bibnamefont{Kun}}
  %\bibinfo{author}{\bibfnamefont{M.~H.} \bibnamefont{Costa}},
  %\bibinfo{author}{\bibfnamefont{R.~N.} \bibnamefont{{Costa Filho}}},
  %\bibinfo{author}{\bibfnamefont{J.~S.} \bibnamefont{{Andrade Jr.}}},
  %\bibinfo{author}{\bibfnamefont{J.~B.} \bibnamefont{Soares}},
  %\bibinfo{author}{\bibfnamefont{S.}~\bibnamefont{Zapperi}}, \bibnamefont{and}
  %\bibinfo{author}{\bibfnamefont{H.~J.} \bibnamefont{Herrmann}},
  \bibinfo{author}{\bibnamefont{\textit{et al.}}},
  \bibinfo{journal}{J. Stat. Mech.} p. \bibinfo{pages}{02003}
  (\bibinfo{year}{2007});
\bibinfo{author}{\bibfnamefont{F.}~\bibnamefont{Kun}}
  \bibinfo{author}{\bibnamefont{\textit{et al.}}},
%   \bibinfo{author}{\bibfnamefont{H.~A.} \bibnamefont{Carmona}},
%   \bibinfo{author}{\bibfnamefont{J.~S.} \bibnamefont{Andrade}},
%   \bibnamefont{and} \bibinfo{author}{\bibfnamefont{H.~J.}
%   \bibnamefont{Herrmann}}, 
  \bibinfo{journal}{Phys. Rev. Lett.}
  \textbf{\bibinfo{volume}{100}}, \bibinfo{pages}{094301}
  (\bibinfo{year}{2008}).

\bibitem[{\citenamefont{Marder}(2000)}]{Marder2000}
\bibinfo{author}{\bibfnamefont{M.~P.} \bibnamefont{Marder}},
  \emph{\bibinfo{title}{Condensed Matter Physics}}
  (\bibinfo{publisher}{Wiley-Interscience}, \bibinfo{year}{2000}).

\bibitem[{\citenamefont{note1}()}]{note1}
\bibinfo{note}{This assumption must clearly break down as the maximum stress intensity
factor approaches the critical value $K_c$ at which the crack becomes unstable, and
which is strongly dependent on the material composition.
In the case of rapid crack growth our approximation
is not valid anymore since inertia effects will play a
dominant role. In this regime, the Paris law itself is no longer verified.}

\bibitem[{\citenamefont{note2}()}]{note2}
\bibinfo{note}{
The damage increment given by Eq. (\ref{eq:df}) intends to
mimick, at the micrometer scale, the various atomistic processes leading to
mechanical fatigue crack growth. Although an analogue of the Paris law
has been observed in samples subject to stress corrosion fatigue,
our approach should not be expected to model the physico-chemical mechanisms at work during corrosion. For a discussion of these points see e.g. 
%Refs. \cite {Lawn1993} and \cite{Ojala2003}.
B. Lawn, \textit{Fracture of brittle solids} (Cambridge University Press, 1993), 2nd ed.,
and also I. O. Ojala \textit{et al.}, J.
Geophys. Res. \textbf{108}, 2268 (2003).
}








\bibitem[{\citenamefont{Barenblatt and Botvina}(1983)}]{Barenblatt1983}
\bibinfo{author}{\bibfnamefont{G.~I.} \bibnamefont{Barenblatt}}
  \bibnamefont{and} \bibinfo{author}{\bibfnamefont{L.~R.}
  \bibnamefont{Botvina}}, \bibinfo{journal}{Izvestiya, USSR Ac. Sci., Mech.
  Solids} \textbf{\bibinfo{volume}{44}}, \bibinfo{pages}{161}
  (\bibinfo{year}{1983}).

\bibitem[{\citenamefont{Sornette and Andersen}(2006)}]{Sornette2006}
\bibinfo{author}{\bibfnamefont{D.}~\bibnamefont{Sornette}} \bibnamefont{and}
  \bibinfo{author}{\bibfnamefont{J.~V.} \bibnamefont{Andersen}},
  \bibinfo{journal}{Europhys. Lett.} \textbf{\bibinfo{volume}{74}},
  \bibinfo{pages}{778} (\bibinfo{year}{2006}).

\bibitem[{\citenamefont{Alava et~al.}(2006)\citenamefont{Alava, Nukala, and
  Zapperi}}]{Alava2006}
\bibinfo{author}{\bibfnamefont{M.~J.} \bibnamefont{Alava}},
  \bibinfo{author}{\bibfnamefont{P.~K. V.~V.} \bibnamefont{Nukala}},
  \bibnamefont{and} \bibinfo{author}{\bibfnamefont{S.}~\bibnamefont{Zapperi}},
  \bibinfo{journal}{Adv. Phys.} \textbf{\bibinfo{volume}{55}},
  \bibinfo{pages}{349} (\bibinfo{year}{2006}).

\bibitem[{\citenamefont{Pradhan et~al.}(2005)\citenamefont{Pradhan, Hansen, and
  Hemmer}}]{Pradhan2005}
\bibinfo{author}{\bibfnamefont{S.}~\bibnamefont{Pradhan}},
  \bibinfo{author}{\bibfnamefont{A.}~\bibnamefont{Hansen}}, \bibnamefont{and}
  \bibinfo{author}{\bibfnamefont{P.~C.} \bibnamefont{Hemmer}},
  \bibinfo{journal}{Phys. Rev. Lett.} \textbf{\bibinfo{volume}{95}},
  \bibinfo{pages}{125501} (\bibinfo{year}{2005}).

% \bibitem[{\citenamefont{Harris}(1974)}]{Harris1974}
% \bibinfo{author}{\bibfnamefont{A.~B.} \bibnamefont{Harris}},
%   \bibinfo{journal}{J. Phys. C} \textbf{\bibinfo{volume}{7}},
%   \bibinfo{pages}{1671} (\bibinfo{year}{1974}).

% \bibitem[{\citenamefont{Lawn}(1993)}]{Lawn1993}
% \bibinfo{author}{\bibfnamefont{B.}~\bibnamefont{Lawn}},
%   \emph{\bibinfo{title}{Fracture of brittle solids}}
%   (\bibinfo{publisher}{Cambridge University Press}, \bibinfo{year}{1993}),
%   \bibinfo{edition}{2nd} ed.
% 
% \bibitem[{\citenamefont{Ojala et~al.}(2003)\citenamefont{Ojala, Ngwenya, Main,
%   and Elphick}}]{Ojala2003}
% \bibinfo{author}{\bibfnamefont{I.~O.} \bibnamefont{Ojala}},
%   \bibinfo{author}{\bibfnamefont{B.~T.} \bibnamefont{Ngwenya}},
%   \bibinfo{author}{\bibfnamefont{I.~G.} \bibnamefont{Main}}, \bibnamefont{and}
%   \bibinfo{author}{\bibfnamefont{S.~C.} \bibnamefont{Elphick}},
%   \bibinfo{journal}{J. Geophys. Res.} \textbf{\bibinfo{volume}{108}},
%   \bibinfo{pages}{2268} (\bibinfo{year}{2003}).

\end{thebibliography}

\end{document}